\documentstyle[pra,epsf,aps,twocolumn]{revtex}

\def\Mvariable#1{\epsilon}
\def\cm#1{}

\begin{document}
\title{{
Critical Exponents without $ \beta $-Function
}}
\author{Hagen Kleinert%
 \thanks{Email: kleinert@physik.fu-berlin.de \hfil \newline URL:
http://www.physik.fu-berlin.de/\~{}kleinert \hfil
%\newline Phone/Fax:  0049 30 8383034
}}
\address{Institut f\"ur Theoretische Physik,\\
Freie Universit\"at Berlin, Arnimallee 14,
14195 Berlin, Germany}
%\pacs{03.20.+i\\ 04.20.Fy\\ 02.40.+m}
%%%%%%%%%%%%
\maketitle
\begin{abstract}
We point out that the recently developed
strong-coupling theory enables us
to calculate the
three main critical exponents $ \nu, \, \eta, \, \omega$,
from the knowledge of only the two renormalization constants
$Z_\phi$  of wave function and $Z_m$ of mass.
The renormalization constant of the coupling strength
is superfluous, and with it
also the $ \beta $-function, the
 crucial quantity of the renormalization group approach to critical
phenomena.
\end{abstract}

%\pacs{03.20.+i\\ 04.20.Fy\\ 02.40.+m}
%\pacs{}
%%%%%%%%%%%%
%\maketitle
%
~\\
\noindent
{\bf 1.}
Some time ago
we have shown
\cite{sc,criteps,alpha}
that there exists a simple way of extracting the
strong-coupling  properties of a $\phi^4$-theory
from perturbation expansions.
In particular, we were able to find the power behavior of
the renormalization constants
in the limit of large couplings, and from this all critical exponents
 of
the system.
By using the known expansion coefficients of the
renormalization constants
in three dimensions up to six loops
we derived extremely accurate
values for the critical exponents.

This is possible because of the fact that $\phi^4$-theory
displays experimentally observed scaling behavior.
This implies that
the perturbation expansions of the critical exponents,
to be
denoted collectively
by $f(g_B)$,
in powers of some dimensionless bare coupling $g_B$
have all a strong-coupling behavior \cite{confl}
of the form
\begin{equation}
f(g_B)=f^*-\frac{{\rm const}}{g_B^{  \omega/ \epsilon}}+\dots~.
\label{appr}\end{equation}
The number $f^*$ is the critical exponent,
and the power $ \omega $ is the
exponent
observable in the approach to scaling.
The parameter
$ \epsilon =4-D$ denotes as usual the deviation of the space dimension from the
naively scale-invariant dimension $4$.

Apart from $ \omega $, there are two independent
critical exponents, for instance $ \nu $ which rules the
divergence
of the coherence length $\xi\propto |T-T_c|^{- \nu }$
as the temperature $T$ approaches
the critical temperature $T_c$, and the exponent $ \gamma $
which does the same thing for the magnetic susceptibility.
The purpose of this paper is to point out that
the strong-coupling theory developed in \cite{sc,criteps}
allows us to calculate all three critical
exponents from
the perturbation expansions of only the two
 renormalization constants
$Z_\phi$  of wave function and $Z_m$ of mass.
There is no need to
go through the hardest
calulation for the renormalization constant of the coupling strength,
and we do not have to
know the famous $ \beta $-function of the renormalization group approach to critical
phenomena, in which
the exponent $ \omega $ is found from the derivative of the $ \beta $-function at its zero.

{\bf 2.} Let us briefly recall the relevant formulas.
Given the first $N+1$ expansion terms of the critical exponents,
$ f_N(g_B)=\sum_{n=0}^N  a_n  g_B  ^n$,
we assume that the strong-coupling behavior
(\ref{appr})
continues systematically
as an inverse powe series in $g_B^{- \omega / \epsilon }$,
 $f_M(g_B)=  \sum_{m=0}^M b_m  (g_B ^{-2/q}) ^m$,
with some finite convergence radius $g_s$  \cite{JK2} (for expamples see
\cite{Interpolation,PI,JK1}).
Then the $N$th approximation to the
value $f^*$ is obtained from the formula
\begin{eqnarray}
&&f_N^* =\mathop{\rm opt}_{\hat{g}_B}\left[
\sum_{j=0}^N a_{j}^{\rm } \hat{g}_B^j
 \sum_{k=0}^{N-j}
      \left( \begin{array}{c}
              - q j/2 \\ k
             \end{array}
      \right)
     (-1)^{k}   \right] ,
\label{coeffb}\end{eqnarray}
where the expression in brackets
has to be optimized in the variational parameter
$\hat g_B$.
The optimum is
the smoothest of the real extrema. If there are none,
the
turning points serve the same purpose.

The derivation  of this expression
is simple:
We replace $g_B$ in $f_N(g_B)$ trivially by
$\bar g_B\equiv g_B/\kappa^q$ with $ \kappa=1$.
Then we rewrite, again trivially, $ \kappa^{-q}$ as $ (K^2+ \kappa^2-K^2)^{-q/2}$
with an arbitrary parameter $K$.
Each term is now expanded in powers of $r=(\kappa^2-K^2)/K^2$
assuming $r$ to be of the order $g_B$.
The we take the limit $ g_B\rightarrow \infty$
at a fixed  ratio
 $\hat g_B\equiv g_B/K^q$,
so that $K\rightarrow \infty$ like $g_B^{1/q}$ and
$r\rightarrow -1$,
yielding  (\ref{coeffb}).
Since the final result to all orders cannot depend on the
arbitrary parameter $K$, we expect the best result to any finite order to
be optimal at an extremal value of $K$, i.e., of $\hat g_B$.

The  strong-coupling approach to the
limiting value
$r= -1+ \kappa^2/K^2
=-1+O(g_B^{-2/q})$
 implies
the leading correction
 to
$f^*_N$ to be
of the order of $g_B^{-2/q} $.
Application of the theory to a function with the
strong-coupling
behavior
(\ref{appr}) requires therefore
setting the parameter
$q$ equal to $2 \epsilon/ \omega$
in formula (\ref{coeffb}).

For $N=2$ and $3$,
the strong coupling limits (\ref{coeffb})
are very simple.
Defining $ \rho\equiv 1+q/2=1+ \epsilon/ \omega$, one has for
$N=2$:
\begin{eqnarray}
f_2^*= \mathop{\rm opt}_{\hat{g}_B}\left[
a_0+
a_1\rho \hat g_B +
a_2\hat g_B^2
\right] =a_0-\frac{1}{4}\frac{a_1^2}{a_2} \rho^2,
\label{f2@}\end{eqnarray}
and for $N=3$:
\begin{eqnarray}
f_3^*&=&  \mathop{\rm opt}_{\hat{g}_B}\left[
a_0+{\scriptstyle\frac{ 1}{ 2}}a_1 \rho( \rho+1)\hat{g}_B+a_2(2 \rho-1)
\hat{g}_B^2
+a_3\hat{g}_B^3 \right]        \nonumber \\
&=&a_0-\frac{1}{3}\frac{\bar a_1\bar a_2}{a_3}\left(1-\frac{2}{3}r
\right)
+\frac{2}{27}\frac{\bar a_2^3}{a_3^2}\left(1-r \right),
\label{f3@}\end{eqnarray}
where
$
r\equiv
 \sqrt{1-3{\bar a_1a_3}/{\bar a_2^2}}$ and
$ \bar a_1 \equiv {\scriptstyle\frac{ 1}{ 2}}a_1 \rho( \rho+1) $
and
$ \bar a_2 \equiv a_2(2 \rho-1) $.
The
positive
square root must be taken
  to connect $g_3^*$ smoothly to
$g_2^*$ for small $g_B$.
If the square root is imaginary, the optimum is given by
the unique turning point, leading once more to (\ref{f3@}), but
with $r=0$.

Before we can apply formula (\ref{coeffb}),
we must find
the
exponent $\omega$
describing the
approach to scaling.
We do this by
studying the
strong-coupling limit
of
the logarithmic derivative
 $s(g_B)\equiv g_Bf'(g_B)/f(g_B)$
of any critical exponent $f(g_B)$, again via formula
(\ref{coeffb}).
Since $f(g_B)$ approaches  a constant $f^*$
like (\ref{appr}),
its
logarithmic derivative $s(g_B)$
must have the same type of behavior
with $s^*=0$.
This equation
determines
$ \omega $.
For an expansion of the critical exponents
up to order $g_B^3$,
it is easy to find explicit results.
Let us denote the
generic expansion
of the exponents for a moment by
$F(g_B)=A_1g_B+A_2g_B^2+\dots~$.
If the expansions start with a nonzero value
at $g_B=0$, this may be subtracted.
The  logarithmic derivative of $F(g_B)$
is
\begin{eqnarray}
s(g_B)&=&1+\hat A_2 g_B+(2\hat A_3-\hat A_2^2)g_B^2
\nonumber \\&&+
(\hat A_2^3-3\hat A_2\hat A_3+3\hat A_4)g_B^3
+\dots~,
\label{omscal2}\end{eqnarray}
where $\hat A_i=A_i/A_1$.
The expansion coefficients on the right-hand sides
serve as coefficients
$a_0,\,a_1\,a_2$
in formulas
(\ref{f2@}) or (\ref{f3@}), with the
left-hand being set equal to zero
to ensure that $s(g_B)\rightarrow s^*=0$
and thus
$ f(g_B)\rightarrow f^*=$const
for $g_B\rightarrow \infty$.

Another way to determine
$ \omega $ uses the fact that
if $F(g_B)$ approaches $F^*$
as in
(\ref{appr}),
the function
\begin{eqnarray}
h(g_B)&\equiv& g_B\frac{F''(g_B)}{F'(g_B)} =2\hat A_2 g_B
+(-4\hat A^2_2+6\hat A_3)g_B^2
\nonumber \\&&
+(8\hat A_2^3-18\hat A_2 \hat A_3 +12\hat A_4)g_B^3
%\nonumber \\ &&+(-16A_2^4+48A_2^2A_3-18A_3^2-32A_2A_4)g_B^4
+\dots~
\label{extraeq}\end{eqnarray}
has the strong-coupling limit
\begin{equation}
h(g_B)\rightarrow h^*=-\frac{ \omega} \epsilon-1.
\label{fome@}\end{equation}

{\bf 3.}
These formulas
can  now be
applied
directly to the power series of the renormalization constants
of mass and wave functions.
Their power behavior for $g_B\rightarrow \infty$
is of the form
\begin{equation}
\frac{m^2}{m_B^2}\propto  g_B^{- \eta_m/ \epsilon}\propto m^{ \eta_m},~~~~~~
\frac{\phi^2}{\phi_B^2}\propto  g_B^{ \eta/ \epsilon}\propto m^{ -  \eta}.
\label{phimass}\end{equation}
The powers functions $ \eta _m$ and $ \eta $ can therefore be calculated
from the strong-coupling limits of the
logarithmic derivatives
\begin{eqnarray}
  \eta_{m}({ g}_B)\!=\!
- \frac{\epsilon~ d}{d\log  g_B}\log\frac{m^2}{m^2_B},~
  \eta({ g_B})
\!=\!
 \frac{\epsilon~ d}{d\log  g_B}\log\frac{\phi^2}{\phi^2_B}
 .\!\!\label{etame2}
\label{etaetam@}\end{eqnarray}
These functions have the
perturbation expansions  up to the order $g_B^3$
\begin{eqnarray}
 &&\eta _m(g_B)=\frac{n+2}{3 }g_B-\frac{n+2}{18}\left(5+2\frac{n+8} \epsilon\right)g_B^2
 \label{@etam}
 \\
&&+\,\frac{2 + n }{108} {
{\left[
       3\left( 37 + 5n \right) \! +   \!
      \frac{ 244 + 38n }{ \epsilon }\!+\!
\frac{ 4\,{{\left( 8 + n \right) }^2}}{ \epsilon ^2}   \right]g_B^3 }
}, \label{@eta}\\
 &&\eta(g_B)=\frac{n+2}{18}g_B^2
-{  \left( 16 + 10\,n + {n^2} \right)  \frac{ \left( \epsilon + 8 \right) \,
     }{216\,\Mvariable{ep}}}g_B^3
 .
\end{eqnarray}
The critical exponent $ \nu $ is obtained from the strong-coupling limit of
the function $ \nu(g_B) =1/[2- \eta _m(g_B)]$, whereas the exponent $ \gamma $
is the same limit of the function $ \gamma (g_B)= \nu(g_B)[2- \eta (g_B)]$.
These functions have the
expansions up tp order $g_B^3$:
{\footnotesize
\begin{eqnarray}
 &&\nu(g_B)=\frac{1}2\!+\!\frac{n\!+\!2}{12}{ g}_B\!+\!
\frac{ n\!+\!2 }{72}\left(n\!-\!3\!-\!2\frac{n\!+\!8}{ \epsilon}\right)
 g_B^2 \nonumber \\
&&+\frac{ 2 \!+\! n }{{432
     {{\Mvariable{ep}}^2}}}} {{\left[ 4{{\left( 8 \!+\! n \right) }^2} \!+\!
       {{\Mvariable{ep}}^2}\left( 95 \!+\! 9n \!+\! {n^2} \right)  \!-\!
       2\Mvariable{ep}\left( \!-\!90 \!+\! n \!+\! 2{n^2} \right)  \right] }g_B^3
,
\label{nu}\\
  &&\gamma(g_B)=1\!+\!\frac{n\!+\!2}{6}{ g}_B\!+\! \frac{ n\!+\!2 }{36}\left(n\!-\!4\!-\!2\frac{n\!+\!8}{ \epsilon}\right)
 g_B^2 \label{gamma} \\
&&+\frac{ 2 \!+\! n}
{
432
     {{\Mvariable{ep}}^2}}}
 {{\left[ 8{{\left( 8 \!+\! n \right) }^2} \!+\!
       4\Mvariable{ep}\left( 106 \!+\! n \!-\! 2{n^2} \right)  \!+\!
       {{\Mvariable{ep}}^2}\left( 194 \!+\! 17n \!+\! 2{n^2} \right)  \right] }
g_B^3.
\nonumber\end{eqnarray}
}
There is no need to give more expansion coefficients,
which can all
be downloaded from the internet  (URL: )

\begin{figure}[tbhp]
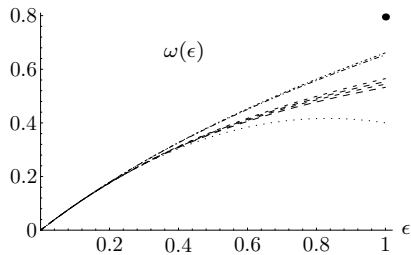

~\\~\\~\\~\\\input omegas.tps
\caption[]{Plots of the various
solutions for $ \omega ( \epsilon )$ obtained
from the perturbation expansions of $ \nu(g_B) $, $ \gamma(g_B) $.
The dotted curve is the universal $ \epsilon $-expansion
up to $ \epsilon ^2$ of
Eq.~(\protect\ref{@epsexp}).
The dashed curves are ordered with increasing dash lengths,
from
(\protect\ref{@gamma2}),
(\protect\ref{@nu2}),
(\protect\ref{@gammar}),
(\protect\ref{@nur}). The dashed-dotted curves
were calculated in Ref.~\protect\cite{criteps} from
the perturbation expansion of $g(g_B)$, once as in
(\protect\ref{@gammar}),
(\protect\ref{@nur}), and once as in
(\protect\ref{@gamma2}),
(\protect\ref{@nu2}).
The dot shows the
accurate value derived below from a seven-loop expansion
in three dimensions.
 }
\label{omegaf}\end{figure}

{\bf 4.} We begin by calculating
the critical exponent $ \omega$
from the
requirement that the expansions
$\nu(g_B)$
has
a constant
strong-coupling limit.
Then also the subtracted function
$F(g_B)= \nu(g_B)-\nu(0)$
has a constant limit.
From (\ref{nu}) and (\ref{omscal2})
we find the expansion coefficients of the logarithmic
derivative
{\footnotesize
\begin{eqnarray}
&&a_0=1,~a_1=\frac{1 }{6}\left(n\!-\!3\!-\!2\frac{n\!+\!8}{ \epsilon}\right),
\label{@} \\
&&a_3 =
\frac{1}{36 \epsilon ^2}\left[{{4{{\left( 8 \!+\! n \right) }^2} \!-\!
     4\Mvariable{ep}\left( \!-\!66 \!-\! 4n \!+\! {n^2} \right)  \!+\!
     {{\Mvariable{ep}}^2}\left( 181 \!+\! 24n \!+\! {n^2} \right) }}\right] .
\nonumber \end{eqnarray}}
Inserted into
(\ref{f2@}), the condition of a zero strong-coupling limit yields
{\footnotesize
\begin{eqnarray}
 \rho =2\frac{{\sqrt{{{4{{\left( 8 \!+\! n \right) }^2} \!-\!
         4\Mvariable{ep}\left( \!-\!66 \!-\! 4n \!+\! {n^2} \right)  \!+\!
         {{\Mvariable{ep}}^2}\left(
 181 \!+\! 24n \!+\! {n^2} \right) }}}}}{
2{{\left( 8 \!+\! n \right) }}  \!+\! \Mvariable{ep}
(n-3)}.
\label{@nur}\end{eqnarray}
}
Let us also derive $ \omega $ from the
vanishing of the logarithmic derivative of the critical exponent $ \gamma $.
Here we find
{\footnotesize
\begin{eqnarray}
 \rho =
2\frac{\sqrt{{{4{{\left( 8 \!+\! n \right) }^2} \!-\!
	 4\Mvariable{ep}\left( \!-\!74 \!-\! 5n \!+\! {n^2} \right)  \!+\!
	 {{\Mvariable{ep}}^2}\left( 178 \!+\! 25n \!+\! {n^2} \right)
 }}}}{
2{{\left( 8 \!+\! n \right) }}  \!+\! \Mvariable{ep}
(n-3)}.
.
\label{@gammar}\end{eqnarray}
}
The resulting exponents $ \omega = \epsilon /( \rho -1)$
are plotted against
 $ \epsilon $ in Fig.~\ref{omegaf}, together with the
 plots derived in Ref.~\cite{criteps}
from the expansion of the renormalized coupling constant
$g(g_B)$ in powers of $g_B$.
The first two terms of the $ \epsilon $-expansion of $ \omega $
are, in all expressions
\begin{eqnarray}
  \omega= \epsilon  -3\frac{3n+14}{(n+8)^2} \epsilon^2+\dots~.
\label{@epsexp}\end{eqnarray}

Let us also calculate $ \omega $ from the
auxiliary function $h(g_B)$ of Eq.~(\ref{extraeq})
by solving Eq.~(\ref{fome@})
for $ \omega $,
which reads more explicitly:
\begin{eqnarray}
-\frac{ \omega}{ \epsilon}-1= -\frac{ \rho}{ \rho-1}&=&
-\frac{1}{2}\frac{\hat A_2^2 \, \rho^2  }{3\hat A_3-2\hat A_2^2}
.
\label{@omega2}\end{eqnarray}
yielding for $ \rho=1+ \epsilon / \omega  $:
\begin{equation}
 \rho=1+ \frac{\epsilon}\omega =\frac{1}{2}+ \sqrt{\frac{6\hat A_3}{\hat A_2^2}-\frac{15}{4}}.
\label{@}\end{equation}
From $F(g_B)= \nu (g_B)-\nu(0)$ we obtain
{\footnotesize
\begin{eqnarray}
 &&\rho =1/2+ \frac{\sqrt{3}}2\label{@nu2}\\
&&\times  \frac{ \sqrt{ 12\,{{\left( 8 + n \right) }^2} -
    12\,\Mvariable{ep}\,\left( -80 - 7\,n + {n^2} \right)  +
    {{\Mvariable{ep}}^2}\,\left( 715 + 102\,n + 3\,{n^2} \right)}}
{2{{\left( 8 \!+\! n \right) }}  \!+\! \Mvariable{ep}
(n-3)},                       \nonumber
\label{@}\end{eqnarray}
}
while   $F(g_B)=  \gamma  (g_B)- \gamma (0)$
yields
{\footnotesize
\begin{eqnarray}
 &&\rho =1/2+ \frac{\sqrt{3}}2\label{@gamma2}\\
&&\times  \frac{ \sqrt{
4\,{{\left( 8 + n \right) }^2} +
    \Mvariable{ep}\,\left( 352 + 32\,n - 4\,{n^2} \right)  +
    {{\Mvariable{ep}}^2}\,\left( 232 + 36\,n + {n^2} \right)
}}
{2{{\left( 8 \!+\! n \right) }}  \!+\! \Mvariable{ep}
(n-3)}.                       \nonumber
\label{@}\end{eqnarray}
}
The resulting $ \omega = \epsilon /( \rho -1)$ are also plotted in
Fig.~\ref{omegaf}. The $ \epsilon $-expansions are again
the same as in (\ref{@epsexp}).

{\bf 5.} With $ \omega $ being determined, the
critical exponents
 $ \nu $ and $ \gamma $ are calculated as before
in Ref.~\cite{criteps}
by inserting their power series coefficients
into the strong-coupling equation
(\ref{f2@}).
Depending on the different expressions
for the resummed $ \omega ( \epsilon )$
we obtain
from the limiting values $f^*$
various resummed functions $ \nu ( \epsilon )$ and $ \gamma ( \epsilon )$.
Their common $ \epsilon $-expansions up to $ \epsilon ^2$ are
\begin{eqnarray}
&& \nu=\frac{1}{2}+\frac{1}{4}\frac{n+2}{n+8} \epsilon+\frac{(n+2)(n+3)(n+20)}{8(n+8)^3} \epsilon^2,
 \label{@nuequa}\\
 && \gamma={1}+\frac{1}{2}\frac{n+2}{n+8} \epsilon
+\frac{1}{4}\frac{(n+2)(n^2+22n+52)}{(n+8)^3} \epsilon^2.
 \label{@gammaequa}
\label{@}\end{eqnarray}

Thus we have shown that the $ \epsilon $-expansions of
all critical exponents including
$ \omega $ can be obtained
from only two renormalization constants
$\phi^2/\phi^2_B$ and $m^2/m_B^2$. Tus
there is no need to calculate
the renormalization constant of the coupling strength.

\begin{figure}[p]
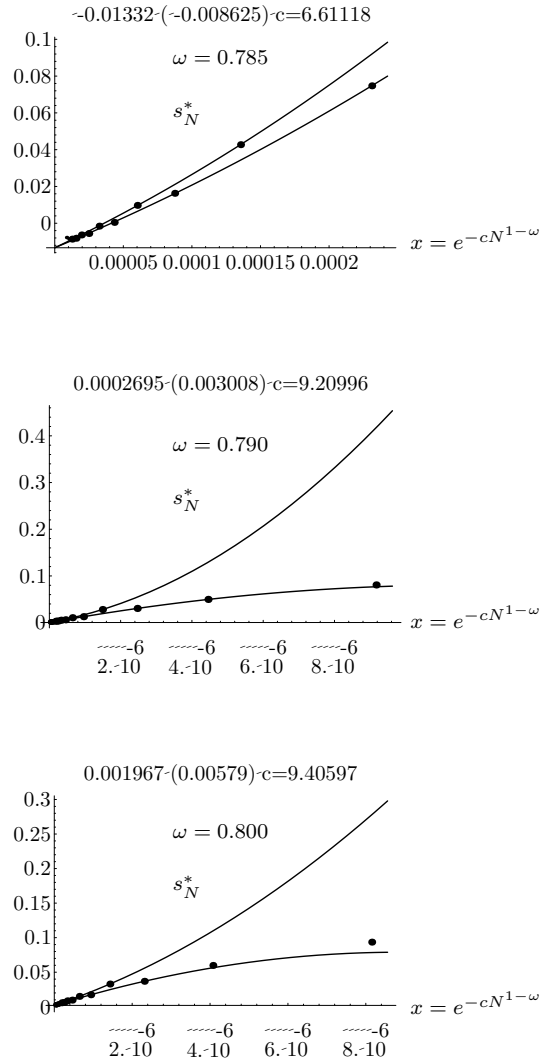

~\\[2cm]
\input extr2.tps\\[3cm]
\input extr.tps\\[3cm]
\input extr3.tps\\
\caption[]{Successive approximations to the strong-coupling
of the logarithmic derivative $s(g_B)=d\log \bar  \eta (g_B)/d\log g_B(g_B)$
and their extrapolation to infinite order for the critical exponents
$ \omega=0.785$,
$ \omega=0.790$, and
$ \omega=0.800$,
showing that the best value is  $ \omega =0.790$. The numbers on top
specify the limiting
value $s^*_\infty$, the last calulated value of order 11,
and the parameter $c$ in
the
theoretically expected large-order
behavior which should be nearly linear in $x=e^{-cN^{1- \omega }}$,
plotted as curves.
The calculated values $s^*_N$ are fitted best
separately for even and odd orders $N$, and the conditions that the two curves intercept at $x=0$
fixes $c$.
 }
\label{@extr}\end{figure}

{\bf 6.}
In three dimensions, the two renormalization constants
$\phi^2/\phi^2_B$ and $m^2/m_B^2$ have been calculated up to seven loops
\cite{MN}, and the principal
critical exponents have been derived
via conventional resummation methods in
 \cite{MN}, \cite{GZ}, and via variational perturbation theory
by the present author  in
\cite{seven,alpha}.
The the latter works,
$ \omega $ was derived from the
expansion of the renormalized coupling constant
in powers of $g_B$, which is only known up to six loops.
In the spirit of the present discussion, we would like to recalculate $ \omega $
from one of the two expansions
of the two renormalization constants known up to seven loops.
As an example we take the expansion for
$\bar \eta \equiv  \eta - \eta _m$. The reson for this choice is
that the theoreticla large-order behavior can be fitted extremely well to the
known  seven expansion coefficients,
so that higher-order coefficients
can be
predicted quite reliably.
This was done in \cite{seven}, and the
extrapolated expansion
can be found
in Table V of that paper.
From this we may determine
$ \omega $ via the condition
that
$s(g_B)=d\log \bar  \eta (g_B)/d\log g_B(g_B)$
vanishes in the strong-coupling limit,
i.e., we the optimum of Eq.~(\ref{coeffb})
for $s^*$ should be zero. We do this only for the universality class of the
superfluid transition of helium,
 proceeding as follows:
For a various $ \omega $-values
we calulate for increasing orders
the strong-coupling values $s^* _N$
and extrapolate them to infinite $N$ by a procedure explained in
Ref.~\cite{seven}.
From the results we find the $ \omega $-value
at which
$s^*_\infty$ is zero to be $ \omega =0.790$.
The extrapolation of $s^*_N$ for this is shown in Fig.~\ref{@extr}

{\bf 7.} The reader may wonder why we can so easily discard
 the renormalized coupling strength $g(g_B)$.
The answer is simple: Instead of $g(g_B)$, we can
just as well parametrize the coupling strength
by the parameter $g_ \nu (g_B)\equiv  [\nu (g_B)- \nu (0)]/ \nu '(0)$
or $g_ \gamma  (g_B)\equiv  [ \gamma  (g_B)-  \gamma  (0)]/ \nu '(0)$
which both start  out like
$$g_ {\nu, \gamma }=g_B-\frac{1}{3 }\left[\frac{n+8}\epsilon +(2-n/2)\right]g_B^2
+O(g_B^3),$$
but continue differently.
The renormalized coupling constant $g(g_B)$ has the same first terms
except for the last parentheses.
For either of these expansions we can define $ \beta $-type of functions
\begin{eqnarray}
 \beta _{ \nu , \gamma }(g_B)=- \epsilon g_{ \nu , \gamma }(g_B)
\frac{d\log g_{ \nu , \gamma }(g_B)
}{d\log g_B},
\label{@}\end{eqnarray}
and express these in terms of
$g_{ \nu , \gamma }$.
Up to the order $ g_{ \nu , \gamma } ^2 $, this yields
\begin{eqnarray}
&& \beta _{ \nu  }(g_{ \nu  })=- \epsilon g_ \nu
+\left[ \frac{8+n}{3}+
\epsilon \frac{3-n}{6}\right] g_ \nu ^2,\nonumber \\
&& \beta _{ \gamma }(g_{  \gamma })=- \epsilon g_  \gamma +
\left[ \frac{8+n}{3}+
\epsilon \frac{2-n}{6}\right] g_  \gamma  ^2.
\label{@}\end{eqnarray}
whose zeros determine the crtical exponents
$ \nu , \gamma $ to satisfy
\begin{eqnarray}
&& g_{ \nu , \gamma }=\frac{3 \epsilon }{8+n}+\dots~,
\label{@}\end{eqnarray}
in agreement with the first two terms in
(\ref{@nuequa}) and (\ref{@gammaequa}). The slopes
of these $ \beta $-like functions
are universally equal to $ \omega $.

{\bf 8.}
Summarizing we see that
no $ \beta $-function is needed to calculate the principal critical exponents
of $\phi^4$-theories,  $ \nu  $, $ \gamma $, amnd $ \omega $,
which can all be obtained from
the two
renormalization constants
of wave function and mass, and thus
from the perturbation expansion of
the self-energy $ \Sigma ({\bf p}^2)$
in powers of the bare coupling constant.
The two desired expansions are extracted
from
 $ \Sigma (0)$ and  $ \Sigma' (0)$.
This observation should  be useful
for possible future automated computer
calculations of the critical exponents.

~~\\~~\\
{\bf Acknowledgment}~\\~\\
The author thanks Dr. A. Pelster
and M. Bachmann for useful discussions.

%******************************************************
%********************************************************************
%
%

%
%
%
%

\pagebreak

\end{document}